\documentclass[aps,prl,onecolumn,showpacs,12pt, superscriptaddress,a4paper]{revtex4-1}
\usepackage{graphicx}
\usepackage{amssymb,amsmath}
\usepackage{epsfig}
\usepackage{txfonts}

\usepackage{amsmath,amssymb,amsfonts}
\usepackage{mathrsfs}
\usepackage{empheq}

\usepackage[utf8]{inputenc}
\usepackage{amsmath}
\usepackage{amsfonts}
\usepackage{amssymb}
\usepackage{dcolumn}   % needed for some tables
\usepackage{multirow}
\usepackage{threeparttable}
\usepackage{setspace}

\usepackage{color}
\usepackage[unicode,breaklinks]{hyperref}

\usepackage[unicode]{hyperref}
\hypersetup{
	pdftitle={Anomalous},
	pdfauthor={Yu, Bradley},
	pdfsubject={Vortex fluids},
	colorlinks=true,
	linkcolor=blue,
	citecolor=blue,
	filecolor=black,
	urlcolor=blue
}
\def\urlprefix{}
\def\url#1{}

\newcommand{\be}{\begin{equation}}
\newcommand{\ee}{\end{equation}}

\let\stdsection\section
\renewcommand\section{\clearpage\stdsection}

\begin{document}
\title{Bandgap Control in Two-Dimensional Semiconductors via Coherent Doping of Plasmonic Hot Electrons}	
\author{Yu-Hui Chen}\thanks{These authors contributed equally}
\affiliation{\scriptsize School of Physics, Beijing Institute of Technology, Beijing 10081,  China} 
\author{Ronnie R. Tamming}\thanks{These authors contributed equally}
\affiliation{\scriptsize MacDiarmid Institute for Advanced Materials and Nanotechnology, Dodd-Walls Centre for Photonic and Quantum Technologies, School of Chemical and Physical Sciences, Victoria University of Wellington, Wellington 6012, New Zealand} 
\author{Kai Chen} 
\affiliation{\scriptsize MacDiarmid Institute for Advanced Materials and Nanotechnology, Dodd-Walls Centre for Photonic and Quantum Technologies, School of Chemical and Physical Sciences, Victoria University of Wellington, Wellington 6012, New Zealand} 
\author{Zhepeng Zhang}
\affiliation{\scriptsize Department of Materials Science and Engineering, College of Engineering, Center for Nanochemistry (CNC), College of Chemistry and Molecular Engineering, Academy for Advanced Interdisciplinary Studies, Peking University, Beijing 100871, China}
\author{Yanfeng Zhang}
\affiliation{\scriptsize Department of Materials Science and Engineering, College of Engineering, Center for Nanochemistry (CNC), College of Chemistry and Molecular Engineering, Academy for Advanced Interdisciplinary Studies, Peking University, Beijing 100871, China}
\author{Justin M. Hodgkiss}
\affiliation{\scriptsize MacDiarmid Institute for Advanced Materials and Nanotechnology, Dodd-Walls Centre for Photonic and Quantum Technologies, School of Chemical and Physical Sciences, Victoria University of Wellington, Wellington 6012, New Zealand} 
\author{Richard J. Blaikie}
\affiliation{\scriptsize MacDiarmid Institute for Advanced Materials and Nanotechnology, Dodd-Walls Centre for Photonic and Quantum Technologies, Department of Physics, University of Otago, PO Box 56, Dunedin 9016, New Zealand} 
\author{Boyang Ding}
\email{boyang.ding@otago.ac.nz}
\affiliation{\scriptsize MacDiarmid Institute for Advanced Materials and Nanotechnology, Dodd-Walls Centre for Photonic and Quantum Technologies, Department of Physics, University of Otago, PO Box 56, Dunedin 9016, New Zealand} 
\author{Min Qiu}
\email{qiu\textunderscore lab@westlake.edu.cn}
\affiliation{\scriptsize Key Laboratory of 3D Micro/Nano Fabrication and Characterization of Zhejiang Province, School of Engineering, Westlake University, 18 Shilongshan Road, Hangzhou 310024, Zhejiang Province, China}
\affiliation{\scriptsize Institute of Advanced Technology, Westlake Institute for Advanced Study, 18 Shilongshan Road, Hangzhou 310024, Zhejiang Province, China}

\date{\today}

\begin{abstract}
	\textbf{Bandgap control is of central importance for semiconductor technologies. The traditional means of control is to dope the lattice chemically, electrically or optically with charge carriers. Here, we demonstrate for the first time a widely tunable bandgap (renormalisation up to 650 meV at room-temperature) in two-dimensional (2D) semiconductors by coherently doping the lattice with plasmonic hot electrons. In particular, we integrate tungsten-disulfide (WS$_2$) monolayers into a self-assembled plasmonic crystal, which enables coherent coupling between semiconductor excitons and plasmon resonances. Accompanying this process, the plasmon-induced hot electrons can repeatedly fill the WS$_2$ conduction band, leading to population inversion and a significant reconstruction in band structures and exciton relaxations. Our findings provide an innovative and effective measure to engineer optical responses of 2D semiconductors, allowing a great flexiblity in design and optimisation of photonic and optoelectronic devices.}
\end{abstract}

\maketitle

Two-dimensional (2D) semiconductors, such as transition metal dichalcogenides (TMDCs)\cite{Mak2010,Splendiani2010}, have direct bandgap at their monolayer limit, exhibiting tremendous potential in development of next-generation nanoscale devices. Like in their bulk counterparts, bandgap control plays a vital role in 2D semiconductor technoglogies, since it enables the creation of desirable optoelectronic properties that are required in numerous applications, ranging from lasers\cite{Ye2015a} to modulators\cite{Mak2016a}, photodetectors\cite{Lopez-Sanchez2013} and photocatalysis\cite{Voiry2013a}. The traditional means of control is to dope the lattice chemically\cite{Kim2015a}, electrically\cite{Chernikov2015d} or optically\cite{Chernikov2015e} with charge carriers, the practicality of which is, however, limited by many factors, e.g. the irreversible bandgap modification, contact-type control and requirement of ultrastrong pump.

Here we report that one can flexibly and effectively modify the electronic band structures of 2D semiconductors by establishing coherent strong coupling between the semiconductor excitons and a plasmonic resonator\cite{Ebbesen1998c,Liu2008}. In particular, plasmonic resonators are metallic nanostructures that support collective oscillation of electrons, known as plasmons. The excitation of plasmons can produce hot electrons, i.e. highly energetic electrons with non-equilibrium thermal distributions\cite{Clavero2014c,Brongersma2015}, which, in the strong coupling regime, can repeatedly dope the lattice along with the coherent plasmon-exciton energy exchange. As a result, the bandgap of 2D semiconductors is significantly renormalised and the renormalisation can be easily altered through changing the detuning between plasmons and excitons.

\begin{figure} [h]
	\includegraphics[width=6.5in]{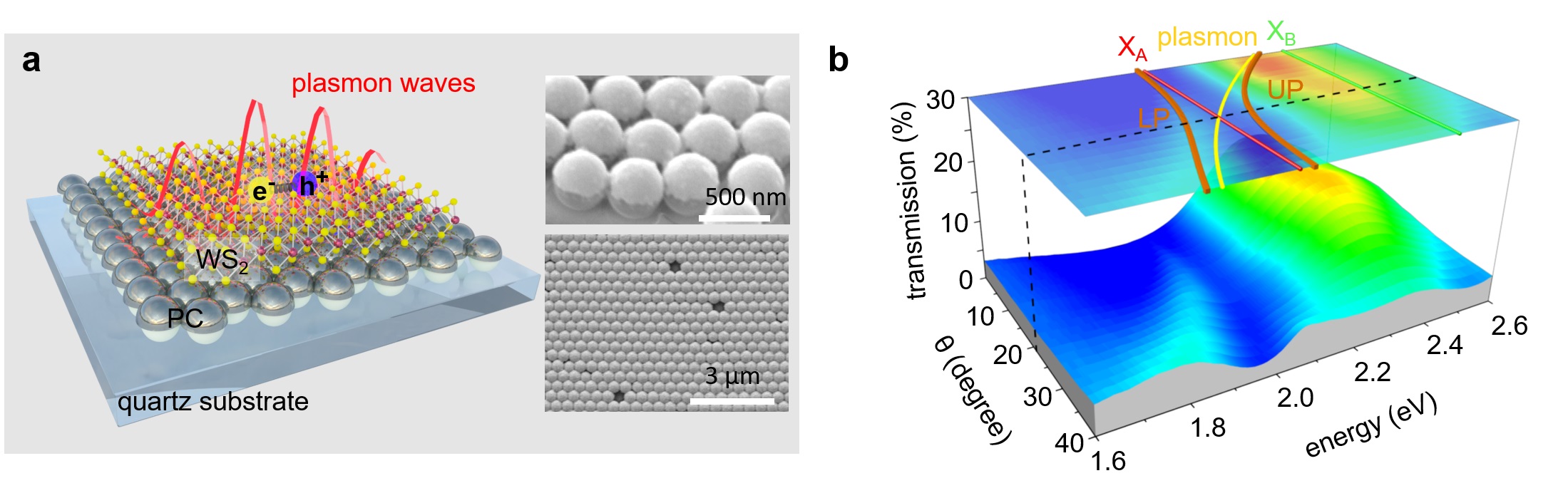}
	\caption{\textbf{Structures of a PC-WS$_\text{2}$ sample and steady-state optical properties.} \textbf{a}, schematic of polariton formation in a WS$_\text{2}$ ML that is supported on a self-assembled plasmonic crystal. The Al$_2$O$_3$ spacer is not depicted for similicity. right insets: side and top-view scanning electron microscope (SEM) images; \textbf{b}, angle-resolved transmission spectra under p-polarised illumination and their projection (top x-y plane), in which the spectral positions of exciton A (X$_\text{A}$) and B (X$_\text{B}$), calculated dispersions of plasmon lattice modes (yellow curve), and upper and lower branches of polaritons (orange curves) are indicated. The tuned angle ($\theta = 22^{\circ}$) is marked with a blacked dahsed line. Refer to Section 1 in the SI for detailed discussion of the strong plasmon-exciton coupling and its dispersion.}
	\label{F1}  
\end{figure}

The schematic of our sample in Fig.\ref{F1}a demonstrates a WS$_2$ monolayer (ML) deposited onto a plasmonic crystal (PC)\cite{Ding2013b,Ding2019}, which comprises of a periodic array of silver capped silica nanospheres that are coated with an ultrathin Al$_2$O$_3$ spacer. This metal-insulator-semiconductor configuration constitutes PC-WS$_2$ hybrid systems, supporting plasmon lattice modes propagating on the PC-WS$_2$ interface. Here the top WS$_2$ MLs belong to the family of atomically thin TMDCs, having been extensively studied\cite{Ye2014a,Sie2017a,Ruppert2017b,Cunningham2017,Steinhoff2017} for their unusual exciton-dominated optical responses, such as high absorption and emission efficiency. These properties make the PC-WS$_2$ systems a suitable platform to study plasmon-exciton interactions\cite{Ding2019}.

The PC geometries were chosen to excite plasmon lattice modes\cite{Ebbesen1998c,Liu2008,Ding2013b,Ding2019} that can match the frequency of exciton A in WS$_\text{2}$ MLs at certain incident angles $\theta$. The plasmon modes show red-shift dispersion at higher $\theta$ (yellow curve in Fig.\ref{F1}b), matching the frequency of exciton A ($E = 2.061$ eV) at $\theta = 22^{\circ}$. In this case, plasmon modes can coherently couple with excitons, leading to the formation of plasmon-exciton polaritons, i.e. half-light half-matter quasiparticles that inherit properties from both the plasmonic and excitonic components. As a result, the transmission maxima exhibit pronounced splitting features that follow the dispersions of upper polariton (UP) and lower polariton (LP), indicating the establishment of strong coupling between plasmons and excitons. When the frequency of the plasmon mode is tuned in resonance with exciton A ($\theta = 22^{\circ}$), the hybrid system is characterised by a vacuum Rabi splitting of $\hbar\cdot\Omega_\text{R}\approx136$ meV.  More detailed analysis of strong plasmon-exciton coupling in equilibrium states can be found in a previous work\cite{Ding2019} and Fig.S1 in the Supplementary Information (SI).

\begin{figure} [t]
	\includegraphics[width=6in]{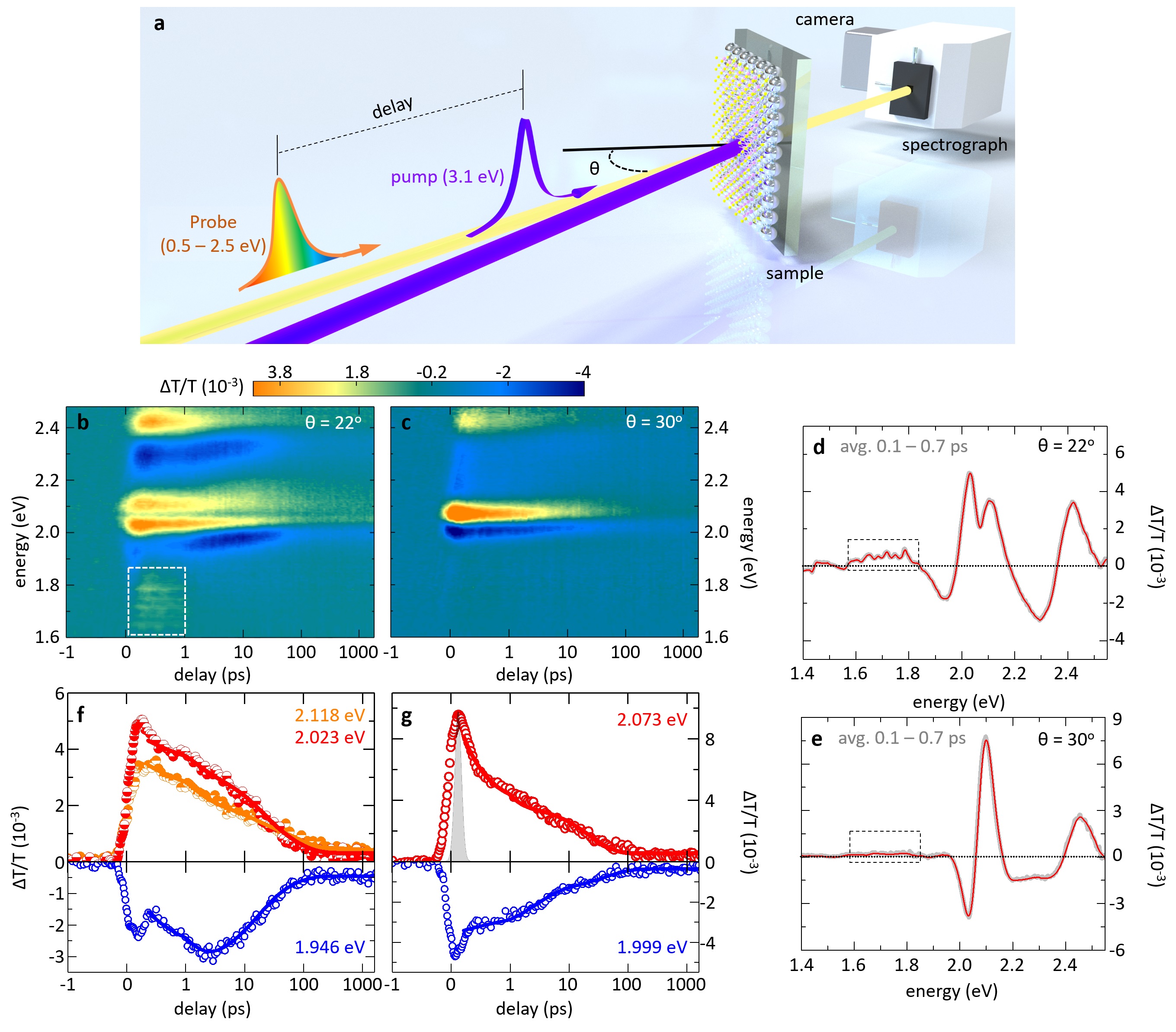}
		\caption{\textbf{Transient optical responses.} \textbf{a}, schematic of angle-resolved ultrafast pump-probe spectroscopy; \textbf{b}, \textbf{d} and \textbf{f} refer to normalised differential transmission spectra ($\Delta \text{T}/\text{T}$) at the tuned angle ($\theta = 22^{\circ}$), while \textbf{c}, \textbf{e} and \textbf{g} refer to $\Delta \text{T}/\text{T}$ at the detuned angle ($\theta = 30^{\circ}$); \textbf{b} and \textbf{c} are intensity plots of $\Delta \text{T}/\text{T}$ as function of time delay and probe photon energy, using the same colour bar (which is also used by Fig.\ref{F3}\textbf{a}); \textbf{d} and \textbf{e} are $\Delta \text{T}/\text{T}$ spectra averaged within the time span from 0.1 to 0.7 ps after pump; \textbf{f} and \textbf{g} are $\Delta \text{T}/\text{T}$ transient at specific energies (labelled with different colours), in which scatter symbols and solid curves represent measured and fitted data, respectively. Dashed frames in panel \textbf{b}, \textbf{d} and \textbf{e} mark the spectral region of the broad maxima (see main text). All measurements were carried out using 400 nm ($E = 3.1$ eV) pump pulses that have 100 fs duration and pump fluence of 12 $\mu$J/cm$^2$ at room temperature. The instrument-response-function is shown as the grey area in panel\,\textbf{g}}
	\label{F2}  
\end{figure}

Upon photoexcitation, the transient optical responses of PC-WS$_2$ samples can be characterised using femtosecond transient absorption (TA) spectroscopy (Fig.\ref{F2}a and Methods), which enables incident angle-resolved probes of the optical properties and dynamics of WS$_2$ MLs that are strongly coupled with plasmon resonances\cite{Darby2015}. Fig.\ref{F2}b shows the transient transmission spectra ($\Delta\text{T}/\text{T}$) with a pump fluence of $12\mu \text{J/cm}^{2}$ as a function of time delay and energy at the tuned state ($\theta = 22^{\circ}$), which displays two split relaxation traces flanking the spectral position of exciton A ($E = 2.061$ eV), corresponding to UP and LP. This sharply contrasts with the single-trace relaxations of exciton B ($E = 2.471$ eV, Fig.\ref{F2}b) and uncoupled exciton A in bare WS$_2$ MLs (Fig.S2 in SI). When the PC is detuned from exciton A, e.g. at $\theta = 30^{\circ}$ (Fig.\ref{F2}c), a single relaxation trace appears, highly resembling the trace of bare exciton A. These ultrashort timescale results confirm again the strong coupling nature of our PC-WS$_2$ systems.

It is worth noting that the photoinduced absorption minimum associated with tuned polaritons  appears at the 1 to 10 ps range (blue area centred at $E = 1.946$ eV in Fig.~\ref{F2}b and the corresponding $\Delta\text{T}/\text{T}$ transient with negative magnitudes in Fig.\ref{F2}f), obviously delayed compared to the minimum near exciton B (Fig.\ref{F2}b) and its counterpart in the detuned polaritons (Fig.~\ref{F2}c), which all emerge simultaneously after the arrival of the pump pulse. Similar postponed minima have been found in transient spectra of bare TMDC MLs, which typically arise from enhanced exciton-exciton and/or exciton-electron interactions under high-power pump that can populate high-density carriers in the lattice\cite{Ceballos2016c,Ruppert2017b,Cunningham2017,Sie2017a} (see Section 2 in SI for detailed discussions). What is different is that, in our hybrid systems, the delayed minima appear under much lower pump intensity than that in the reference experiments for bare WS$_2$ MLs and are only associated with tuned polaritons. 

More importantly, it is noted that a $\Delta\text{T}/\text{T}$ maximum lasting for $\sim1$ ps in $E = 1.6$ to $1.8$ eV arise in the tuned polariton spectra (Fig.\ref{F2}b), which, in contrast, is remarkably  weaker in the detuned state (Fig.\ref{F2}c) and is completely absent in bare WS$_2$ MLs (Fig.S2 in SI). The integrated $\Delta\text{T}/\text{T}$ spectrum near zero probe delay (Fig.\ref{F2}d) shows that the broad maximum has positive magnitudes, which indicates negative optical absorption or positive gain, being a clear evidence of bandgap renormalisation accompanied by population inversion.\cite{Chernikov2015e} Such phenomena are typically induced by the population of high-density carriers in 2D semiconductor lattice\cite{Meckbach2018}, which leads to the non-equilibrium occupation of electron and/or hole states that can induce the formation of new quasiparticle bandgaps. This process can be decribed by\cite{Peyghambarian1993}:

\begin{equation}
\label{BG}
\Delta E_\text{g} = -\underset{q\neq 0}{\sum}V_\text{s}(q)\,[f_\text{e}(q)+f_\text{h}(q)]-\underset{q\neq 0}{\sum}[V_\text{s}(q)-V(q)]
\end{equation}

\noindent where $V_\text{s}(q)$ and $V(q)$ represent fourier transforms of screened and unscreened Coulomb potentials, while $f_\text{e}(q)$ and $f_\text{h}(q)$ are occupation probabilities of electron and hole with momentum $q$. The onset of the new bandgap can be extracted from the low-energy end of the broad maximum. It means that in our experiments, the renormalised bandgap starts at $E_\text{g} \approx 1.60$ eV, lying $\sim400$ meV below LP and $\sim650$ meV below the initial bandgap of WS$_2$ MLs (given that the binding energy of exciton A is $\sim200$ meV\cite{Cunningham2017}). This is, to the best of our knowledge, the largest bandgap renormalisation in 2D semiconductors under such a low pump intensity (12$\mu$J$/$cm$^2$) to date, which, in the meanwhile, results in the inversion of carrier population near the newly formed band edge\cite{Chernikov2015e,Meckbach2018}, presenting as optical gains, i.e. the broad maximum in Fig.\ref{F2}b and \ref{F2}d.

These unusual spectral and transient features are broadly understood as the presence of high-density carriers,  which, in our tuned PC-WS$_2$ systems, surprisingly have been achieved under room temperature and extremely low pump intensity ($12\mu$J$/$cm$^2$). This sharply contrasts with similar observations\cite{Chernikov2015e} in bare WS$_2$ single/bi-layers with ultrastrong photoexcitation ($840\mu$J$/$cm$^2$ at 70 K or $3400\mu$J$/$cm$^2$ at room temperature). In their study, the population of high-density carriers are a result of Mott-transition, which are induced by enhanced exciton-exciton interactions under high-power pump, reducing exciton binding energy, finally breaking excitons into unbound electron-hole plasma\cite{Chernikov2015e,Steinhoff2017}. In our experiments, the pump power is too low to develop a Mott-transition, suggesting that there must be other sources that can provide large numbers of additional carriers. 

To understand the origin of these carriers, we turn to discuss one unique property of plasmon-exciton polaritons, i.e. the generation of hot electrons that are inherited from the polaritons' plasmon root. In particular, hot electrons are electrons with non-equilibrium thermal distributions, generated by plasmon dephasing from wave-like states through non-radiative decay\cite{Brongersma2015}, which can electrically dope adjacent semiconductors\cite{Fang2012}, modifying their photovoltaic and photocatalytic performance\cite{Clavero2014c}. When plasmons are coupled to exciton-like resonances in semiconductors, the hot electron density can be highly enhanced in the lattice through direct electron tunneling\cite{GarciaDeArquer2013} or dipole-dipole interaction\cite{Cushing2012}.  Therefore it is very likely that the high-density carriers in tuned PC-WS$_2$ systems are the hot electrons introduced during strong coupling process. (See Section 3 in SI for detailed discussions)

The analyses of relaxation dynamics of tuned and detuned polaritons support the hot electron model. We note that both the UP and LP in Fig.\ref{F2}f demonstrate slower decays than that of detuned states in Fig.\ref{F2}g  (Table S2 in SI for fitting parameters). This observation coincides with a previous study\cite{Boulesbaa2016d}, clearly indicating the involvement of plasmonic hot electrons in the strong plasmon-exciton coupling process. Specifically, as the system sits in the strong coupilng regime, after photoexcitation, excitons and plasmons coherently exchange energy at the Rabi frequency ($\sim136$ meV)\cite{Vasa2013}, while the plasmon-to-exciton process is accompanied by hot electron population in the lattice. Such a charge population runs at an ultrashort period of $\sim30$ fs ($T_\text{R}=2\pi/\Omega_\text{R}$), which is too short to be caught by our equipment, also greatly shorter than the exciton formation ($<1$ ps)\cite{Ceballos2016c}, the non-radiative decay (at scales of $10$ ps) and the radiative decay process (up to few-hundred ps) in WS$_2$ MLs\cite{Ruppert2017b,Sie2017a}. This means that during exciton relaxation, there is frequent tunneling/generation of hot electrons that can repeatedly fill the unoccupied states in conduction band of WS$_2$ monolayers, which slow down the exciton bleaching via Pauli blocking and lead to the extended lifetimes (Section 4 in SI for more details). 

Given that there is little evidence for other possible carrier sources, e.g.  polariton condensates \cite{Byrnes2014}, we conclude that coherent doping of plasmonic hot electrons is the origin of the spectral and transient features that require high-density population. In particular, the hot electron population repeatedly takes place throughout the whole relaxation process, while the Al$_2$O$_3$ spacer can form a Schottky-like barrier that prevents charges from returning back to the metals\cite{Cushing2012,GarciaDeArquer2013}. As a result, hot electrons can be accumulated in the lattice before they decay (within 1 ps\cite{Brongersma2015}), which simultaneously competes with rapid exciton relaxations, transiently converting the intrinsic WS$_2$ monolayers to "\textit{n}-doped" ones. This leads to the giant bandgap renormalisation with population inversion that peak at few-hundred femtoseconds (Fig.S10 in SI), and also induces the delayed absorption minima in Fig.\ref{F2}b and \ref{F2}f (Section 5 in SI).

\begin{figure} [!ht]
	\includegraphics[width=6in]{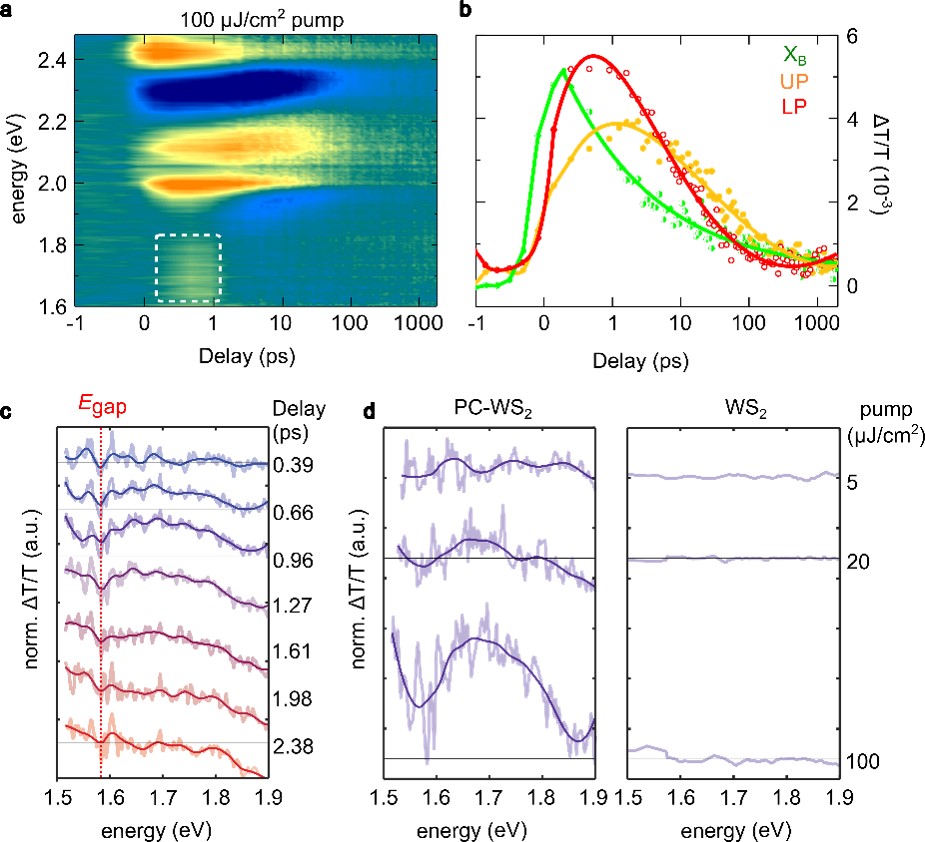}
	\caption{\textbf{Bandgap renormalisation and evloution of population inversion.} \textbf{a}, intensity plot of $\Delta\text{T}/\text{T}$ spectra of PC-WS$_2$ under $100\mu$J$/$cm$^2$ pump fluence at $\theta = 22^{\circ}$, where orange (blue) colour represents the maximum (minimum) value. \textbf{b}, delay time dependent spectra ($\Delta\text{T}/\text{T}$) at energies of UP, LP and exciton B extracted from panel \textbf{a}. Solid curves are plotted only for visual guidance \textbf{c},\,$\Delta\text{T}/\text{T}$ spectra at different delay times, extracted from the white dashed frame in panel \textbf{a}; red dashed vertical line indicates the onset of renormalised bandgap. \textbf{d}, comparison of $\Delta\text{T}/\text{T}$ spectra at delay of 0.96 ps between PC-WS$_2$ (left) and WS$_2$ MLs (right) under gradually increasing pump fluence.}
	\label{F3}  
\end{figure}

To confirm our observations, we have performed meaurements under $\sim10$ times higher pump fluence ($100\mu$J$/$cm$^2$) (Fig.~\ref{F3}a). Apart from the pronounced broad maxima at low energies, we can see large spectral shift as well as remarkably delayed occurance of UP and LP maxima, revealing that the accumulation of hot electrons competes with the relaxation dynamics, which significantly enhances the systems' nonlinear responses on ultrashort timescales. (Detailed discussions in Section 6 of SI). Similar to the low-power case, the transient variation of the broad maximum (Fig.~\ref{F3}c) under intense photoexcitation takes $\sim 1.5$ ps from initial excitation to fading. Fig.~\ref{F3}d shows the evolution of population inversion, where the magnitude and width of the maximum is highly dependent on pump intensity. Under $100\mu$J$/$cm$^2$ pump fluence, the full-width at half-maximum can reach at $\sim200$ meV with highly enhanced magnitudes as compared to the maximum under $5\mu$J$/$cm$^2$ pump, also contrasting the unchanged flat spectral features in bare WS$_2$ MLs. But, even in this case, the pump fluence is still sigfinicantly lower than that in Ref.\cite{Chernikov2015e}.

As discussed above, the strong plasmon-exction coupling dramatically modifies the electronic band structures of WS$_2$ monolayers, which are induced, to a large degree, by plasmonic hot electron doping via strong coupling. This effect is extremely hard to observe in traditional exciton-polaritons\cite{Byrnes2014}, being a non-trivial factor that has to be considered when studying light-matter interactions using plasmonic resonators, which, on the other hand, provides new and effective measures to engineer bandgap of 2D semiconductors.

\clearpage

\section*{Acknowledgments} 
The authors acknowledge the New Idea Research Funding 2018 (Dodd-Walls Centre for photonic and quantum technologies), the Marsden Fast-start Fund by Royal Society of New Zealand through contract MFP-UOO1827 and the Smart Ideas Fund by Ministry of Business, Innovation and Employment, New Zealand through contract UOOX1802. In addition, this work was supported in part by the National Key Research and Development Program of China (no. 2017YFA0205700) and the National Natural Science Foundation of China (nos. 61425023, 61235007, 61575177 and 51861135201). The authors also acknowledge the visiting Fellowship awarded by New Zealand Centre at Peking University. We thank Dr. M. Yan and Dr. F. Hong for their help with thin-film deposition, AFM, and SEM measurements.

\section*{Author Contributions} 
B.D. and Y.-H.C. conceived the project; Z.Z, and B.D. prepared the samples; R.T., K.C., Y.-H.C. and B.D. carried out the optical and other characterization; Y.-H.C. and B.D. performed the simulation;
Y.Z., M.Q., R.J.B., and B.D. supervised the projects; Y.-H.C. and B.D. prepared the manuscript; all authors discussed and analyzed the results.

\bibliography{UF_Polaritons}

\clearpage

\end{document}